\documentclass[conference]{IEEEtran}

\ifCLASSINFOpdf
\else
\fi
\usepackage[cmex10]{amsmath}
\usepackage{amssymb}

\newtheorem{theorem}{Theorem}

\newtheorem{corollary}{Corollary}
\newtheorem{definition}{Definition}
\newtheorem{example}{Example}
\newtheorem{remark}{Remark}

\newcommand{\beq}[1]{\begin{equation}\label{#1}}
\newcommand{\eeq}{\end{equation}}
\newcommand{\req}[1]{(\ref{#1})}
\newcommand{\sopr}[1]{\widetilde {\overline{#1}}}

\newcommand{\eq}{\triangleq}

\newcommand{\x}{{\textbf{\textit{x}}}}
\newcommand{\y}{{\textbf{\textit{y}}}}

 \newcommand{\1}{{\bf 1}}

\newcommand{\A}{{\cal A}}
\newcommand{\An}{{\cal A}^n}
\newcommand{\p}{{\sf p}}

\newcommand{\Mc}{{\cal M}}

\newcommand{\X}{{\bf X}}
\newcommand{\D}{{\cal D}}

\renewcommand{\a}{{\bf a}}
\renewcommand{\b}{{\bf b}}

\renewcommand{\S}{{\cal S}}

\begin{document}

\title{On Critical Relative Distance of DNA Codes
for Additive  Stem Similarity}

\author{\IEEEauthorblockN{A.~D'yachkov, A.~Voronina}
\IEEEauthorblockA{Department of Probability Theory,\\
 Faculty of Mechanics and Mathematics,\\
 Moscow State University,\\Moscow, 119992, Russia,\\
 Email: agd-msu@yandex.ru,\\vorronina@gmail.com.}
\and
\IEEEauthorblockN{A.~Macula, T.~Renz}
\IEEEauthorblockA{Air Force Res. Lab.,\\
 IFTC, Rome Research Site,\\Rome NY 13441, USA,\\
 Email: macula@geneseo.edu,\\thomas.renz@rl.af.mil.}
\and
\IEEEauthorblockN{and V.~Rykov}
\IEEEauthorblockA{Department of Mathematics,\\
 University of  Nebraska at Omaha,\\
 6001 Dodge St., Omaha,\\NE 68182-0243, USA,\\
 E-mail: vrykov@mail.unomaha.edu.}}


\maketitle

\begin{abstract}

We consider DNA codes based on the nearest-neighbor (stem) similarity model
which adequately reflects the "hybridization potential" of two DNA sequences.
Our aim is to present a survey of bounds  on the rate of DNA codes with respect to a thermodynamically
motivated similarity measure called an additive stem similarity.
These results yield a method to analyze and compare known samples  of the nearest neighbor
"thermodynamic weights"   associated to stacked pairs that occurred in DNA
secondary structures.

\end{abstract}



\section{Introduction}

 Single strands of DNA are represented by oriented sequences with elements
 from alphabet $\A\eq\{A,C,G,T\}$.
The \textit{reverse-complement}
(Watson-Crick transformation) of a DNA
strand is defined by first reversing the order of the letters and then
substituting each letter $x$ for its complement $\bar{x}$,
namely: $A$ for $T,$ $C$ for $G$ and
vice-versa. For example, the reverse complement of \textit{AACG} is
\textit{CGTT}.
For strand $\x=(x_1x_2\dots x_{n-1}x_n)\in\A^n=\{A,C,G,T\}^n$, let
\beq{sop}
\sopr{\x}=(\bar{x}_n\bar{x}_{n-1}\dots\bar{x}_{2}\bar{x}_{1})\in\A^n=
\{A,C,G,T\}^n
\eeq
denote its  reverse  complement.
If $\y=\sopr{\x}$, then $\x=\sopr{\y}$ for any~$\x\in \A^n$.
If $\x=\sopr{\x}$, then $\x$ is called a {\em self reverse
complementary} sequence. If $\x\ne\sopr{\x}$, then a pair $(\x\,,\,\sopr{\x})$
is called a {\em pair of mutually reverse complementary}
sequences. A (perfect) \textit{Watson-Crick duplex}
is the joining of oppositely directed $\x$ and $\sopr{\x}$ so that every
letter  of one strand is paired with its complementary
letter  on the other strand
in the double helix structure, i.e., $\x$ and $\sopr{\x}$
are "perfectly compatible." However, when two, not
necessarily complementary, oppositely directed DNA strands are "sufficiently
compatible," they too are capable of coalescing into a double stranded DNA
duplex. The process of forming DNA duplexes from single strands is referred to
as \textit{DNA hybridization. Crosshybridization} occurs when two oppositely
directed and non-complementary DNA strands form a duplex.

 In general, crosshybridization is undesirable as it usually leads to experimental error.
 To increase the accuracy and throughput of the applications listed in~\cite{bres86}-\cite{san04},
 there is a desire to have collections of DNA strands, as large and as mutually
 incompatible as possible, so that no crosshybridization can take place.
 It is straightforward to view this problem as one of coding theory~\cite{ms77}.

 DNA nanotechnology often requires collections of DNA strands called \textit{free energy gap codes}~\cite{Dyachkov07}
 that will correctly "self-assemble" into Watson-Crick duplexes and do not produce erroneous crosshybridizations.
 When these collections consist entirely of pairs of mutually reverse complementary DNA strands
 they are called \emph{DNA tag-antitag systems}~\cite{kad03} and \emph{DNA codes}~\cite{Dyachkov07}-\cite{Dyachkov05}.

 The best known to date \emph{biological} model, which is commonly utilized to estimate hybridization energy is the "nearest-neighbor"{}
 similarity model introduced in~\cite{bres86}. Roughly, it implies that hybridization energy for any two DNA strands should be
 calculated as a sum of \emph{thermodynamic weights} of all \emph{stems} that were formed in the process of hybridization.
 Stem is defined as a pair of consecutive DNA letters of either of the strands, which coalesced with a pair of consecutive
 DNA letters of the other DNA strand. This biological model leads to a special \emph{similarity function} on the space $\An$.

 First known to authors constructions of DNA codes were suggested in~\cite{Rykov00}-\cite{Marathe}. They were based on conventional Hamming
 distance  codes. Some methods of combinatorial coding theory have been developed~\cite{Milenkovic}-\cite{Abualrub}
 as a means by which such DNA codes  can be
 found.  From the very beginning it was understood that
 hybridization energy for DNA strands should be somehow simulated with the similarity function for sequences from $\An$.
 But it can be easily noticed, that Hamming similarity does not in the proper degree inherit
 the idea of "nearest-neighbor"{} similarity model. Thus there is no wonder that further exploration
 activities primarily focused on the search of appropriate similarity function.

 One example of such function was proposed in~\cite{Dyachkov00b}, where it was calculated
 as the sum of weights of all elements,
 constituting the longest common Hamming subsequence. Later attempts
 included deletion similarity~\cite{dvtw00},
 which was earlier introduced by Levenshtein~\cite{Levenshtein01}
 and block similarity~\cite{d05a}-\cite{Dyachkov05}. Both functions are non-additive
 which allowed for consideration of such cases as shifts of DNA sequences along each other.
 Nevertheless, all of them still did not catch the point of "nearest-neighbor"{} similarity model.

 In 2008 we published our first work~\cite{dv08}, devoted to the study of \emph{stem} similarity functions.
 There we considered the simplest case, when similarity between two sequences from $\An$ is equal to the number
 of stems  in the longest common Hamming subsequence between these two sequences.
 The common stem  is understood as a block of length 2
 which contains two adjacent elements of both of the initial sequences.


 In~\cite{dv09},  we introduced the concept of an additive stem $w$-similarity for
 an arbitrary weight function $w=w(a,b)>0$,  defined
 for all 16 elements  $(ab)\in\A^2$, called stems.
 To calculate the additive stem $w$-similarity between two DNA sequences one
 should add up weights of all stems in the longest common Hamming subsequence
 between them (see, below Definition~1).
 Finally, our  recent  works~\cite{d08}-\cite{dv10} deal with non-additive stem $w$-similarity function,
 previously introduced in~\cite{Dyachkov07}. The given model also implies counting the weights of all formed stems between
 two DNA sequences with only difference that these stems are contained not in Hamming common subsequence but in subsequence
 in sense of Levenstein insertion-deletion metric. To find more detailed discussion of applicability of proposed
 constructions for modeling DNA hybridization assays please refer to work~\cite{Dyachkov07}.

 In current report we will summarize  main results of~\cite{dv09} in study of asymptotic behavior of DNA codes maximal size
 for  additive  stem $w$-similarity function. We will show how these results lead to the development
 of possible criteria called a {\em critical relative $w$-distance of DNA codes}
 for distinguishing between  weight samples $w(a,b)$ found in different experiments. We will also explain,
 how our consideration prompts the algorithms for composing DNA ensembles of optimal
 size for the given length of DNA strands.

 \section{Additive Stem $w$-Similarity Model}

 \subsection{Notations and Definitions}

The symbol $\eq$ denotes definitional equalities
 and the symbol $[n]\eq\{1,2,\dots,n\}$ denotes the set of
 integers from 1 to~$n$.
  Let $w=w(a,b)>0$, $\,a,b\in\A$, be a weight function such that
 \beq{w}
 w(a,b)=w(\bar{b},\bar{a}), \quad a,b\in \A.
 \eeq
 Condition~\req{w} means that $w(a,b)$ is an
 invariant function under Watson-Crick transformation.

 \begin{definition}~\cite{Dyachkov07},\cite{dv09}.\quad
 For $\x,\y\in\A^n$, the number
 \beq{sim-w}
 \S_{w}(\x,\y)\,\eq\,\sum\limits_{i=1}^{n-1}\,s^{w}_i(\x,\y),\quad\mbox{where}\quad
 $$
 $$
 s^{w}_i(\x,\y)\,\eq\,
 \begin{cases}
 w(a,b) &  \mbox{if}\,\, x_i=y_i=a,\; x_{i+1}=y_{i+1}=b,\\
 0      &  \mbox{otherwise},%
 \end{cases}
 \eeq
 is called an {\em additive stem $w$-similarity} between $\x$ and~$\y$.
 \end{definition}

Function $\S_w(\x,\sopr{\y})$
is used to model a {\em thermodynamic similarity} ({\em hybridization
energy}) between DNA sequences $\x$ and~$\y$.
In virtue of~\req{w}-\req{sim-w}
the function
\beq{Sw1}
\S_w(\x,\y)\,=\,\S_w(\y,\x)\,\le\,\S_w(\x,\x),\quad \x,\y\,\in \A^n
\eeq
In addition,
\beq{Sw2}
\S_w(\x,\sopr{\y})\,=\,\S_w(\y,\sopr{\x}),
\qquad \x,\y\,\in \A^n.
\eeq
 Identity~\req{Sw2}  implies
the symmetry property of hybridization energy
between DNA sequences $\x$ and~$\y$~\cite{Dyachkov07}-\cite{Dyachkov05}.

 \begin{example}
 In~\cite{dv08} we considered  {\em constant weights} $w=w(a,b)\equiv1$,
 $a,b\in\A$, for which the additive stem $1$-similarity
$\S_{\1}(\x,\y)$, $0\le\S_{\1}(\x,\y)\le\S_1(\x,\x)= n-1$, is the above-mentioned number
 of \emph{stems} in the longest common Hamming subsequence between $\x$ and~$\y$.
 \end{example}

 \begin{example}
Table~1  shows a biologically motivated collection of  weights
$w(a,b)\eq U(a,b)$ called~\cite{san98} {\em unified weights}:
\begin{center}
\begin{tabular}{|c||c|c|c|c|}
\hline
$U(a,b)$ & $b=A$  & $b=C$  &  $b=G$  & $b=T$ \\
\hline
\hline
$a=A$ & 1.00  & 1.44 & 1.28 & 0.88\\
\hline
$a=C$ & 1.45  & 1.84 & 2.17 & 1.28\\
\hline
$a=G$ & 1.30  & 2.24 & 1.84 & 1.44\\
\hline
$a=T$ & 0.58  & 1.30 & 1.45 & 1.00\\
\hline
\hline
\end{tabular}.
\end{center}
\centerline{Table 1:  Unified weights $U(a,b)$, 1998.}
 The given values $U(a,b)$ are based on weight samples
which  come from~\cite{san98} and~\cite{san04} and are the nearest neighbor
"thermodynamic weights" (e.g., free energy of formation)  associated to
stacked pairs that occurred in DNA secondary structures. See~\cite{zuk99} for an
introduction to the nearest neighbor model.
\end{example}

Taking into account inequality~\req{Sw1}, we give

 \begin{definition}~\cite{Dyachkov07},\cite{dv09}.\quad
 The number
  \beq{dist}
  \D_{w}(\x,\y)\,\eq\,\S_{w}(\x,\x)\,-\, \S_{w}(\x,\y)=
 \sum\limits_{i=1}^{n-1}\,\eta^{w}_i(\x,\y),
 $$
 $$
\eta^{w}_i(\x,\y)\eq  s^{w}_i(\x,\x)- s^{w}_i(\x,\y)\ge0,
 \eeq
 is called an additive stem {\em $w$-distance} between~$\x,\y\in\A^n$.
 \end{definition}

 Let 
 $\x(j)\eq(x_1(j)x_2(j)\dots x_n(j))\in\A^n$,
 $j\in[N]$, be {\em codewords} of a  $q$-ary {\em code}~$\X=\{\x(1),\x(2),\dots,\x(N)\}$
 of {\em length} $n$ and {\em size} $N$, where $N=2,4,\dots$ is an {\em even} number.
Let $\,D$, $\,0\,<\,D\,\le\,\max\limits_{\x\in\A^n}\,\S_w(\x,\x)$,
be an arbitrary positive number.

 \begin{definition}~\cite{Dyachkov07},\cite{dv09}.\quad
 A code $\X$ is called a DNA code of distance~$D$ for additive stem $w$-similarity~\req{sim-w}
 (or a {\em $(n,D)_w$-code}) if the following two conditions are fulfilled.
 $(i)$.~For any integer $j\in[N]$, there exists $j'\in[N]$, $j'\ne j$, such
 that $\x(j')=\sopr{\x(j)}\ne\x(j)$. In other words, $X$ is a collection
 of $N/2$ pairs of mutually reverse complementary sequences.
 $(ii)$.~The minimal $w$-distance of code $X$ is
 \beq{deff}
 \D_w(X)\,\eq\,\min\limits_{j\ne j'}\,\D_{w}\left(\x(j),\x(j')\right)\ge D.
 \eeq
 \end{definition}

 Let $N_{w}(n,D)$ be the {\em maximal} size of DNA $(n,D)_{w}$-codes for distance~\req{dist}.
 If $d>0$ is a fixed number, then
 \beq{R_w}
 R_{w}(d)\,\eq\,\varlimsup\limits_{n\to\infty}\;
 \frac{\log_4\,N_{w}(n,nd)}{n},\qquad d>0,
 \eeq
 is called a {\em rate} of DNA $(n,nd)_w$-codes for the {\em relative distance}~$d>0$.

\subsection{Construction}

 \begin{theorem} \quad
 If $n=2t+1$, $\,t=1,2,\dots$, then
 $$
 N_{\1}(n,n-1)=16.
 $$
 \end{theorem}
 \medskip

 \begin{proof}\quad
 Codewords of $(n,n-1)_{\1}$-code should not contain any common stems with each other.
 Note, that $|\A^2|=16$ and hence for any $(n,n-1)_{\1}$-code $\X=\{\x(1),\dots\x(N)\}$
 $$
 \left|\,\{(x_1(u)x_2(u)),\,\,\, u\in[N]\}\,\right|
 \,\leq\,|\A^2|\,=\,16.
 $$
 Thus,
 $$
 N_{\1}(n,n-1)\,\leq\,16.
 $$

 Obviously, for odd $n$, the set $\An$ doesn't contain self reverse
 complementary words. For  stem $\a=(a_1a_2)\in\A^2$, define
 $\x(\a)=(a_1a_2a_1a_2\dots a_2a_1a_2a_1)\in\An$. Code
 $$
 \X_{r}=\{\x(\a),\,\,\a\in\A^2\}, \quad
 |\X_r|=4^2=16
 $$
 constitute a DNA $(n,n-1)_{\1}$-code of size $16$ for additive stem $1$-similarity. Theorem~1 is proved.
 \end{proof}
 \smallskip

 \begin{example}
 For instance, if $\,n=5$, $\,D=n-1=4\,$, then $8$ pairs of mutually reverse
 complementary codewords of code $\X_r$ are:
 $$
 (AAAAA,\;TTTTT),\quad
 (ACACA,\;TGTGT),
 $$
 $$
 (CCCCC,\;GGGGG),\quad
 (CACAC,\;GTGTG),
 $$
 $$
 (AGAGA,\;TCTCT),\quad
 (ATATA,\;TATAT),
 $$
 $$
 (CGCGC,\;GCGCG),\quad
 (CTCTC,\;GAGAG).
 $$
 \end{example}
 \smallskip

 \begin{remark}
 Note that for any weight function $w$,
the additive stem $w$-similarity
$\S_{w}\left(\x(\a),\x(\b)\right)=0$,
$\a,\b\in\A^2$, $\a\neq \b$.
 Hence, the minimal $w$-distance~\req{deff} of code $\X_r$ is
 $$
 \D_w(\X_r)=\min\limits_{j}\,\S_{w}\left(\x(j),\x(j)\right)\,
 \ge\,2t\cdot\underline{w},
 $$
 where $\underline{w}=\min\limits_{a,b\in\A}w(a,b)$.
 Thus, for  any weight function $w$, the code $\X_r$ is also
 a $(n,(n-1)\cdot\underline{w})_w$-code. For example,
 for the additive stem $U$-similarity of Example~2,
 the number $\D_U(\X_r)=2t$. Therefore, the code $\X_r$ is a $(n,n-1)_U$-code.
 \end{remark}

 \subsection{Bounds on Rate $R_{w}(d)$}

 Let $\p\eq\{\,p(a,b),\;a,b\in\A\}$ be an arbitrary joint probability distribution
 on the set of stems $(ab)\in\A^2$, i.e.,
 $$
 \sum\limits_{a,b\in\A}p(a,b)=1,\qquad  p(a,b)\geq 0
 \quad\mbox{for any}\quad a,b\in\A.
 $$

 To describe bounds on the rate $R_{w}(d)$, we will consider joint probability
 distributions $\p$, such that the corresponding marginal probabilities coincide, i.e., for any $a\in\A$
 \beq{max1}
 p_1(a)\eq
 \sum\limits_{b\in\A}p(a,b)\,=\,\sum\limits_{b\in\A}p(b,a)\,\eq
 \,p_2(a)>0
 \eeq
 and, in addition, function $ p(a,b)$, as well as  weight function~\req{w}, is invariant under
 Watson-Crick transformation, i.e.,
 \beq{max2}
 p(a,b)=p(\overline{b},\overline{a})\quad\mbox{for any}\quad a,b\in\A.
 \eeq
Let
 $$
 p_{1}(b|a)\,\eq\,\frac{p(a,b)}{p_1(a)},\quad
 p_{2}(b|a)\,\eq\,\frac{p(b,a)}{p_2(a)}
 $$
 denote the corresponding  {\em conditional} probabilities.
It is easy to check, that  for distributions $\p$ with properties~\req{max1}-\req{max2},
 and for the corresponding conditional probabilities, the following equalities hold true for any $a,\,b\in\A$:
 \beq{max3}
 p_1(a)=p_2(a)=p_1(\overline{a})=p_2(\overline{a}),\quad
 p_{1}(b|a)=p_{2}(\overline{b}|\overline{a}).
 \eeq

 For a fixed weight function~\req{w}, introduce values
 \beq{T}
 T_w\,\eq\,\max\limits_{\req{max1}}\,T_w(\p),
 $$
 $$
 T_w(\p)\,\eq\,\sum\limits_{a,b\in\A}\,
 \left(p(a,b)-p^2(a,b)\right)w(a,b),
 \eeq
 where the maximum is taken over all distributions $\p$ for which condition~\req{max1} hold true.
 Note, that if weight function is invariant under Watson-Crick transformation, then maximizing
 distribution of~\req{T} will satisfy conditions~\req{max2}-\req{max3}.
\smallskip

Applying an analog of the conventional Plotkin bound~\cite{ms77}, one can prove

 \begin{theorem} \cite{dv09}\quad
 $\,$ If $d\ge T_w$, then $R_w(d)=0$.
  \end{theorem}
\smallskip

Let $\x=(x_1x_2\dots x_n)\in\A^n$ be the stationary  Markov chain
with initial distribution $p_1(a)$, $a\in\A$, and transition matrix
$P=\|p_1(b|a)\|$, $a,b\in\A$, i.e.
\beq{markov}
\Pr\{x_i=a\}\eq p_1(a), \; \Pr\{x_{i+1}=b|x_i=a\}\eq p_1(b|a)\,
\eeq
for any $a,b\in\A$ and $i\in[n-1]$.

 Let a distribution $\p$ satisfy~\req{max1} and let also
 the following \emph{Markov condition} $\Mc$ be fulfilled: {\em transition
 matrix $P$  must define such Markov chain $\x=(x_1x_2\dots x_n)$, that for any
 pair of states $a,b\in\A$ there exists an integer $m\in[4]$ such
 that the conditional  probability $\Pr\{x_{m+1}=b|x_1=a\}>0$}.

 \begin{theorem} \cite{dv09}\quad
 For any probability distribution $\p$, satisfying condition~$\req{max1}$
 and Markov condition $\Mc$, and any relative distance $d$,  $0<\,d\,< T_w(\p)$,
 the rate~$R_w(d)>0$.
  \end{theorem}
\smallskip

Theorem 2 is established using the  ensemble of random codes
where independent codewords $\x=(x_1x_2\dots x_n)$ are identically
distributed in accordance  with the
Markov chain~\req{markov} and, in virtue of~\req{max3}, the
corresponding reverse complement codewords
$\sopr{\x}=(\bar{x}_n\bar{x}_{n-1}\dots\bar{x}_{2}\bar{x}_{1})$
have the same distribution~\req{markov} as well.
In addition, the proof of Theorem 2 is  based on
the Perron-Frobenius theorem (see~\cite{dz93}, Theorem~3.1.1).

 Let $T_w(\p)$ be defined by ~\req{T} and
  \beq{T_M_def}
 T^{\Mc}_w\,\eq\,\max\limits_{\req{max1},\,\Mc}\,T_w(\p).
 \eeq

 If  $\,T_w\,=\,T_w^{\Mc}\,$, then the corresponding weight
 function $w=w(a,b)$
 is called \emph{regular}, and {non-regular} otherwise. If a weight
 function $w=w(a,b)$ is regular, then $T_w$
 is called the \emph{critical relative distance}  of
 $(n,dn)_w$-codes.

 From Theorem~2 and~3 it follows

 \begin{corollary} \cite{dv09}\quad  If a weight
 function $w=w(a,b)$ is regular, then the maximal size of $(n,nd)_w$-codes
 increases exponentially  with increasing $n$ if and only if~$0<d<T_w$.
 \end{corollary}
 \smallskip

 \begin{remark}
 Results of Theorem~2 prompts an idea, that the construction of optimal random DNA codes for additive stem $w$-similarity should be based on generation of independent Markov chains with transition matrix $P$ and initial distribution $p_1(a)$, such that corresponding distribution $\p$ affords maximum in~\req{T_M_def}.
 \end{remark}
 \medskip

\section{Weight Sample Analysis Based on
Criterion\\ of Critical Relative Distance}

In this section, we will discuss  {\em samples of
weight function} (or, briefly, {\em weight samples})  $w=w(a,b)$, $a,b\in\A$,
taken from SantaLucia (1998) (see Table~1 in~\cite{san98}).
In   Tables~2-8, we present  weights $w(A,A)=w(T,T)$ and
samples of  {\em relative} weights $\widetilde{w}(a,b)$ with respect
to $w(A,A)$, i.e., for any $a,b\in\A$,
\beq{rel-w}
\widetilde{w}=\widetilde{w}(a,b)\eq\frac{w(a,b)}{w(A,A)}
,\quad
\widetilde{w}(a,b)=\widetilde{w}(\bar{b},\bar{a}).
\eeq
Pure numbers $\widetilde{w}(a,b)$ are comfortable  for a mutual comparison
and for the comparison with  unified weights of  Table~1.

\begin{center}
\begin{tabular}{|c||c|c|c|c|}
\hline
 $w(A,A)=0.43$ & $b=A$  & $b=C$  &  $b=G$  & $b=T$ \\
\hline
\hline
 $a=A$ & $1.00$  & $2.28$ &  ${\bf1.93} $ &  $0.63$\\
\hline
 $a=C$ & $2.32$  & $2.84$ & $3.95$ & ${\bf1.93}$\\
\hline
 $a=G$ & $2.16$  & $3.81$ & $2.84$ & $2.28$\\
\hline
 $a=T$ & $0.51$  & $2.16$ & $2.32$ & $1.00$\\
\hline
\hline
\end{tabular}\quad
\end{center}
\centerline{Table~2: Gotoh, 1981.}

\begin{center}
\begin{tabular}{|c||c|c|c|c|}
\hline
 $w(A,A)=0.89$ & $b=A$  & $b=C$  &  $b=G$  & $b=T$ \\
\hline
\hline
 $a=A$ & $1.00$  & ${\bf1.35}$ &  $1.52 $ &  $0.91$\\
\hline
 $a=C$ & $1.54$  & $1.84$ & $2.24$ & $1.52$\\
\hline
 $a=G$ & $1.40$  & $2.20$ & $1.84$ & ${\bf1.35}$\\
\hline
 $a=T$ & $0.85$  & $1.40$ & $1.54$ & $1.00$\\
\hline
\hline
\end{tabular}
\end{center}
\centerline{ Table~3: Vologodskii, 1984.}

\begin{center}
\begin{tabular}{|c||c|c|c|c|}
\hline
 $w(A,A)=0.67$  & $b=A$  & $b=C$  &  $b=G$  & $b=T$ \\
\hline
\hline
 $a=A$ & $1.00$  & $1.69$ &  $1.75 $ &  $0.93$\\
\hline
 $a=C$ & $1.78$  & $2.31$ & $2.79$ & $1.75$\\
\hline
 $a=G$ & ${\bf1.67}$  & $2.76$ & $2.31$ & $1.69$\\
\hline
$a=T$ & $1.04$  & ${\bf1.67}$ & $1.78$ & $1.00$\\
\hline
\hline
\end{tabular}\quad
\end{center}
\centerline{ Table~4: Blake, 1991.}

\begin{center}
\begin{tabular}{|c||c|c|c|c|}
\hline
$w(A,A)=0.93$ & $b=A$  & $b=C$  &  $b=G$  & $b=T$ \\
\hline
\hline
$a=A$ & ${\bf1.00}$  & $1.63$ &  $1.11$ &  $0.89$\\
\hline
$a=C$ & $1.35$  & $1.80$ & $1.77$ & $1.11$\\
\hline
 $a=G$ & $1.68$  & $2.62$ & $1.80$ & $1.63$\\
\hline
 $a=T$ & $0.75$  & $1.68$ & $1.35$ & ${\bf1.00}$\\
\hline
\hline
\end{tabular}\quad
\end{center}
\centerline{ Table~5: Benight, 1992.}

\begin{center}
\begin{tabular}{|c||c|c|c|c|}
\hline
 $w(A,A)=1.02$ & $b=A$  & $b=C$  &  $b=G$  & $b=T$ \\
\hline
\hline
 $a=A$ & ${\bf1.00}$  & $1.40$ &  $1.14 $ &  $0.72$\\
\hline
 $a=C$ & $1.35$  & $1.74$ & $2.05$ & $1.14$\\
\hline
 $a=G$ & $1.43$  & $2.24$ & $1.74$ & $1.40$\\
\hline
 $a=T$ & $0.59$  & $1.43$ & $1.35$ & ${\bf1.00}$\\
\hline
\hline
\end{tabular}\quad
\end{center}
\centerline{ Table~6: SantaLucia, 1996.}

\begin{center}
\begin{tabular}{|c||c|c|c|c|}
\hline
  $w(A,A)=1.20$  & $b=A$  & $b=C$  &  $b=G$  & $b=T$ \\
\hline
\hline
 $a=A$ & $1.00$  & ${\bf1.25}$ &  ${\bf1.25} $ &  $0.75$\\
\hline
 $a=C$ & $1.42$  & $1.75$ & $2.33$ & ${\bf1.25}$\\
\hline
 $a=G$ & ${\bf1.25}$  & $1.92$ & $1.75$ & ${\bf1.25}$\\
\hline
 $a=T$ & $0.75$  & ${\bf1.25}$ & $1.42$ & $1.00$\\
\hline
\hline
\end{tabular}\quad
\end{center}
\centerline{ Table~7: Sugimoto, 1996.}

\begin{center}
\begin{tabular}{|c||c|c|c|c|}
\hline
  $w(A,A)=1.66$  & $b=A$  & $b=C$  &  $b=G$  & $b=T$ \\
\hline
\hline
$a=A$ & $1.00$  & ${\bf0.68}$ &  $0.81 $ &  $0.72$\\
\hline
$a=C$ & $1.08$  & $1.66$ & $1.98$ & $0.81$ \\
\hline
$a=G$ & $0.85$  & $1.70$ & $1.66$ & ${\bf0.68}$\\
\hline
$a=T$ & $0.46$  & $0.85$ & $1.08$ & $1.00$\\
\hline
\hline
\end{tabular}\quad
\end{center}
\centerline{ Table~8: Breslauer, 1986.}

\subsection{Analysis of Tables 1-8 for Additive $\widetilde{w}$-Distance}

{\em Analysis of Table~1 and Tables~3-7:}\quad
The given weight samples are regular and the maximum
in~\req{T} is attained
when $p(a,b)=0$ if stem $(ab)\in L_4$, where the set $L_4$ of forbidden
stems in the Markov chain~\req{markov} maximizing~\req{T} has the form
\beq{L4}
L_4\eq\{(AT),(TA),(AA),(TT)\}.
\eeq
Below, in Table~1' and Tables~3'-7', we present the estimated  values of
joint probabilities $p(a,b)$ and marginal probabilities $p_1(a)$
for which the  maximum in~\req{T} is
attained. Values of the critical relative
distance~$T_{\widetilde{w}}$ are given as well.

\begin{center}
\begin{tabular}{|c||c|c|c|c||c|}
\hline
$p(a,b)$ & $b=A$  & $b=C$  &  $b=G$  & $b=T$ & $p_1(a)$  \\
\hline
\hline
$a=A$ & $0$  & $.0589$ &  $.0081 $ &  $0$ &     $.067$   \\
\hline
$a=C$ & $.0610$  & $.1544$ & $.2095$ & $.0081$ & $.433$  \\
\hline
$a=G$ & $.0060$  & $.2136$ & $.1544$ & $.0589$ &  $.433$  \\
\hline
$a=T$ & $0 $  & $.0060$ & $.0610$ & $0$ &   $.067$   \\
\hline
\hline
\end{tabular}\quad
\end{center}
\centerline{ Table~1': Unified weights $U(a,b)$. \quad $T_{U}=1.58$.}

\begin{center}
\begin{tabular}{|c||c|c|c|c||c|}
\hline
$p(a,b)$ & $b=A$  & $b=C$  &  $b=G$  & $b=T$ & $p_1(a)$  \\
\hline
\hline
$a=A$ & $0$  & $.0706$ &  $.0080 $ &  $0$ &     $.078$  \\
\hline
$a=C$ & $.0638$  & $.1411$ & $.2087$ & $.0080$ & $.422$  \\
\hline
$a=G$ & $.0147$  & $.1951$ & $.1411$ & $.0706$ &  $.422$   \\
\hline
$a=T$ & $0 $  & $.0147$ & $.0638$ & $0$ &   $.078$   \\
\hline
\hline
\end{tabular}\quad
\end{center}
\centerline{ Table~3': Vologodskii, 1984. \quad $\;T_{\widetilde{w}}=1.61$.}

\begin{center}
\begin{tabular}{|c||c|c|c|c||c|}
\hline
$p(a,b)$ & $b=A$  & $b=C$  &  $b=G$  & $b=T$ & $p_1(a)$ \\
\hline
\hline
$a=A$ & $0$  & $.0331$ &  $.0346 $ &  $0$ &     $.068$  \\
\hline
$a=C$ & $.0406$  & $.1535$ & $.2037$ & $.0346$ & $.432$ \\
\hline
$a=G$ & $.0270$  & $.2188$ & $.1535$ & $.0331$ &  $.432$ \\
\hline
$a=T$ & $0 $  & $.0270$ & $.0406$ & $0$ &   $.068$   \\
\hline
\hline
\end{tabular}\quad
\end{center}
\centerline{ Table~4': Blake, 1991. \quad $\;T_{\widetilde{w}}=1.97$.}

\begin{center}
\begin{tabular}{|c||c|c|c|c||c|}
\hline
$p(a,b)$ & $b=A$  & $b=C$  &  $b=G$  & $b=T$ & $p_1(a)$  \\
\hline
\hline
$a=A$ & $0$  & $.0675$ &  $.0144$ &  $0$ &     $.082$ \\
\hline
$a=C$ & $.0478$  & $.1326$ & $.2234$ & $.0144$ & $.418$ \\
\hline
$a=G$ & $.0340$  & $.1841$ & $.1326$ & $.0675$ &  $.418$ \\
\hline
$a=T$ & $0 $  & $.0340$ & $.0478$ & $0$ &   $.082$   \\
\hline
\hline
\end{tabular}\quad
\end{center}
\centerline{ Table~5': Benight, 1992. \quad $\;T_{\widetilde{w}}=1.58$.}

\begin{center}
\begin{tabular}{|c||c|c|c|c||c|}
\hline
$p(a,b)$ & $b=A$  & $b=C$  &  $b=G$  & $b=T$ & $p_1(a)$ \\
\hline
\hline
$a=A$ & $0$  & $.0608$ &  $.0095 $ &  $0$ &     $.070$  \\
\hline
$a=C$ & $.0616$  & $.1499$ & $.2087$ & $.0095$ & $.430$  \\
\hline
$a=G$ & $.0087$  & $.2102$ & $.1499$ & $.0608$ &  $.430$  \\
\hline
$a=T$ & $0 $  & $.0087$ & $.0616$ & $0$ &   $.070$  \\
\hline
\hline
\end{tabular}\quad
\end{center}
\centerline{ Table~6': SantaLucia, 1996. \quad $\;T_{\widetilde{w}}=1.55$.}

\begin{center}
\begin{tabular}{|c||c|c|c|c||c|}
\hline
$p(a,b)$ & $b=A$  & $b=C$  &  $b=G$  & $b=T$ & $p_1(a)$ \\
\hline
\hline
$a=A$ & $0$  & $.0507$ &  $.0140 $ &  $0$ &     $.065$  \\
\hline
$a=C$ & $.0444$  & $.1551$ & $.2217$ & $.0140$ & $.435$  \\
\hline
$a=G$ & $.0203$  & $.2091$ & $.1551$ & $.0507$ &  $.435$ \\
\hline
$a=T$ & $0 $  & $.0203$ & $.0444$ & $0$ &   $.065$   \\
\hline
\hline
\end{tabular}\quad
\end{center}
\centerline{Table~7': Sugimoto, 1996. \quad $\;T_{\widetilde{w}}=1.50$.}
\bigskip

{\em Analysis of Table~2:}\quad
The given weight sample is regular and the maximum
in~\req{T} is attained
when $p(a,b)=0$ if stem $(ab)\in L_6$,
where the set $L_6$ of forbidden
stems in the Markov chain~\req{markov} maximizing~\req{T} has the form
\beq{L6}
L_6=\{(AT),(TA),(AA),(TT),(AG),(CT)\}.
\eeq
Below, in Table~2', we present the estimated values of
joint  $p(a,b)$ and marginal $p_1(a)$ probabilities
for which the maximum in~\req{T} is
attained. The estimated  value of critical relative
distances~$T_{\widetilde{w}}=2.60$ is given as well.

\begin{center}
\begin{tabular}{|c||c|c|c|c||c|}
\hline
$p(a,b)$ & $b=A$  & $b=C$  &  $b=G$  & $b=T$ & $p_1(a)$ \\
\hline
\hline
$a=A$ & $0$  & $.0593$ &  $ 0 $ &  $0$ &     $.059$  \\
\hline
$a=C$ & $.0466$  & $.1427$ & $.2515$ & $0$ & $.441$  \\
\hline
$a=G$ & $.0127$  & $.2261$ & $.1427$ & $.0593$ &  $.441$ \\
\hline
$a=T$ & $0 $  & $.0127$ & $.0466$ & $0$ &   $.059$ \\
\hline
\hline
\end{tabular}\quad
\end{center}
\centerline{ Table~2': Gotoh, 1981. \quad $\;T_{\widetilde{w}}=2.60$.}
\bigskip

{\em Analysis of Table~8:}\quad The given weight sample
$\widetilde{w}$ is a
non-regular weight sample because  the maximum in~\req{T} is
attained (with the  maximal value $T_{\widetilde{w}}=1.70$) for probability
distribution $p'(a,b)$, $\,(ab)\in\A^2$, which does not satisfy
\emph{Markov condition} $\Mc$ and  has the form:
\begin{center}
\begin{tabular}{|c||c|c|c|c||c|}
\hline
$p'(a,b)$ & $b=A$  & $b=C$  &  $b=G$  & $b=T$ & $p'_1(a)$  \\
\hline
\hline
$a=A$ & $.0344$  & $0$ &  $ 0 $ &  $0$ &     $.034$   \\
\hline
$a=C$ & $0$  & $.2190$ & $.2466$ & $0$ & $.466$   \\
\hline
$a=G$ & $0$  & $.2466$ & $.2190$ & $ 0$ &  $.466$  \\
\hline
$a=T$ & $0 $  & $0$ & $0$ & $.0344$ &   $.034$  \\
\hline
\hline
\end{tabular}\quad
\end{center}
\centerline{ Table~8': Breslauer, 1986. \quad $\;T'_{\widetilde{w}}=1.70$.}
This implies that for  weight sample $\widetilde{w}$ from Table~8,
we cannot estimate the  critical relative distance   of optimal DNA codes
based on  additive stem $\widetilde{w}$-similarity.

\subsection{Conclusion}
For  regular weight samples from Tables~2-7 (T2-T7),
the descriptive analysis and comparison of critical parameters  are
summarized as follows:
\begin{center}
\begin{tabular}{|c||c||c|c|c|c|c|}
\hline
& T2 & T3 & T4 & T5 & T6 & T7  \\
\hline
\hline
$L$  & $L_6$  & $L_4$ & $L_4$ & $L_4$ & $L_4$ & $L_4$       \\
\hline
$T_{\widetilde{w}}$ & $2.60$  & $1.61$ & $1.97$ & $1.58$ & $1.55$ & $1.50$    \\
\hline
\hline
\end{tabular}\,,
\end{center}
where the corresponding set $L$ ($L=L_4$ or $L=L_6$) of forbidden
stems in  codewords of  optimal DNA codes, for which the
critical relative distance $T_{\widetilde{w}}$ can be attained,
is defined by~\req{L4} or by~\req{L6}.

\end{document}